\documentclass[aps,prl,reprint,superscriptaddress,amsmath,amsfonts]{revtex4-2}
\usepackage{bbm}
\usepackage{graphicx}
\usepackage{mathtools}
\usepackage{xcolor}

\usepackage{braket}
  
\usepackage{lipsum}

\usepackage{verbatim}

\usepackage[showboxes]{textpos}
\usepackage{flafter}
\usepackage{bm}

\usepackage[colorlinks=true, linkcolor=blue, citecolor=blue, urlcolor=blue]{hyperref}

\newcommand{\ketbra}[2]{\ket{#1}\bra{#2}} % for Dirac ketbras

\newcommand{\abs}[1]{\left\lvert {#1} \right\rvert} % absolute value
\newcommand{\op}[1]{\hat{#1}} %  operator
\begin{document}

%\preprint{}

\title{Engineering of spontaneous emission in free space via conditional measurements 
%of a three-atom system
}


%\thanks{}


\author{Manuel Bojer}
\email{manuelbojer6@gmail.com}
\affiliation{Institut f\"ur Optik, Information und Photonik, Universit\"at Erlangen-N\"urnberg, 91058 Erlangen, Germany}
%\affiliation{International Max Planck Research School for the Science of Light (IMPRS-PL), 91058 Erlangen, Germany}

\author{Lukas G\"otzend\"orfer}
\affiliation{Institut f\"ur Optik, Information und Photonik, Universit\"at Erlangen-N\"urnberg, 91058 Erlangen, Germany}
%\affiliation{International Max Planck Research School for the Science of Light (IMPRS-PL), 91058 Erlangen, Germany}

\author{Romain Bachelard}
\affiliation{Universit\'e C\^ote d'Azur, CNRS, Institut de Physique de Nice, 06560 Valbonne, France}
\affiliation{Departamento de F\'isica, Universidade Federal de S\~ao Carlos,
	Rodovia Washington Lu\'is, km 235—SP-310, 13565-905 S\~ao Carlos, SP, Brazil}

%\author{Kevin G\"unthner}
%\affiliation{Quantum Information Processing Group (QIV), Max Planck Institute for the Science of Light (MPL), 91058 Erlangen, Germany}
%\affiliation{Institut f\"ur Optik, Information und Photonik, Universit\"at Erlangen-N\"urnberg, 91058 Erlangen, Germany}

%\author{J\"org Evers}
%\affiliation{Max Planck Institute for Nuclear Physics, Saupfercheckweg 1, 69117 Heidelberg, Germany}

\author{Joachim von Zanthier}
\affiliation{Institut f\"ur Optik, Information und Photonik, Universit\"at Erlangen-N\"urnberg, 91058 Erlangen, Germany}
%\affiliation{International Max Planck Research School for the Science of Light (IMPRS-PL), 91058 Erlangen, Germany}

%\affiliation{Erlangen Graduate School in Advanced Optical Technologies (SAOT), Universit\"at Erlangen-N\"urnberg, 91052 Erlangen, Germany}

\date{\today}

\begin{abstract}
	%As is well-known in Cavity-QED the spontaneous emission properties of an ensemble of atoms can be modified by changing the environment surrounding the atoms. Another possibility of modifying the spontaneous emission is by inducing correlations between the atoms. 
	%We study the emission properties of three identical two-level atoms initially prepared in the excited state by measuring Glauber's third-order intensity correlation function in the far field. We assume that two of the three atoms are close to each other, so that they are subject to the dipole-dipole interaction, while the third one is many wavelenghths far away. Although the third atom does not interact with the first two atoms, its presence changes the emission profile of the last emitted photon dramatically, such that both spatial and temporal modifications arise. This is due to the fact that the detection of the first two photons prepares the system in an entangled state via the measurement process such that the emission of the last excitation is shared by all three atoms. As a consequence, super- or subradiant emission behavior with effective symmetric and antisymmetric decay rates is obtained depending on the emission direction. In this way, modification of the spontaneous emission rates is achieved by adding a third non-interacting atom. 
	\noindent
	We study the collective spontaneous emission of three identical two-level atoms initially prepared in the excited states by measuring Glauber's third-order photon correlation function. Assuming two atoms at sub-wavelength distance from each other such that they are subject to the dipole-dipole interaction while the third one is located  several wavelengths away, we observe super- and subradiant decay alike, depending on the direction of observation. % The behavior results from the fact that the detection of the first two photons prepares the system in an entangled state such that the last emission is shared by all three atoms, and produces a fringe pattern. 
	%This shows that, even though the third atom does not interact with the first two atoms, its presence changes the emission profile of the last emitted photon drastically, producing both spatial and temporal modifications.
	%In this way, via three-photon interference, spatial and temporal modifications of the spontaneous emission rate are achieved via presence of a third atom, even though the latter does not interact with any of the other particles.
	Unlike the case where no remote atom is introduced or no conditional measurements are performed, the spontaneous emission behavior of the conditioned three-atom system differs  strongly from 
	%both in space and time 
	the single-atom and the canonical two-atom configuration. %The conditional measurements associated with the  three-photon correlation function entangle the remote atom with the two other emitters leading to spatial interference whereas the dipole-dipole interaction between the adjacent atoms modifies the decay rate.
	The conditional measurements associated with the  three-photon correlation function in combination with the dipole-dipole interaction between the adjacent atoms lead to quantum interference among the different decay channels allowing to engineer the spontaneous emission in space and time.

\end{abstract}


\maketitle


%\section{Introduction}

\noindent
The interaction between an excited atom and the fluctuations of the electromagnetic vacuum field leads to deexcitation of the atom, a random process called spontaneous emission. A plethora of publications has discussed the possibilities to modify this fundamental process, e.g., by  changing the properties of the vacuum field by use of cavities~\cite{purcell1946, klepppner1981, agarwal1975, Goy1983, eschner2001}, by exploiting nano-optical devices \cite{zoller2017,chang2018colloquium,Lukin2016,Lukin2018,Waks2018,Gammon2019,Rauschenbeutel2020}, or applying external coherent fields~\cite{shiyao1996, paspalakis1998}. Interestingly, an ensemble of interacting emitters in free space also leads to  modified spontaneous emission, e.g., mediated by the dipole-dipole interaction~\cite{lehmberg1970}, giving rise to collective spontaneous emission coined super- and subradiance~\cite{Dicke1954,Agarwal1974,gross1982,ficek2002}. 

The superradiant cascade from the fully-excited state of an atomic ensemble to its ground state, where the system passes by the set of symmetric Dicke states with different decay rates, has been  studied extensively in the past~\cite{Agarwal1974,gross1982}. Recently, there has been renewed interest in the collective emission of atomic ensembles also in the single-excitation regime (sometimes labelled ``single-photon'' regime), where  steady-state shifts~\cite{Ido2005,Scully2009,Keaveney2012,Javanainen2014,Meir2014,Jennewein2016,Bromley2016}, as well as superradiant~\cite{Araujo2016,Roof2016,Blatt2018,Richter2022} and subradiant~\cite{Guerin2016,Das2020,Blatt2018,Richter2022} emission have been reported. The subradiant emission has been observed in particular in the late-time regime, after decoherence has eliminated the short-lived states and populated essentially the long-lived levels~\cite{Cipris2021,Ferioli2021,Santos2021}.
These works have in common that the radiation is typically monitored in a given direction, that is, all the photons are recorded within the same emission angle. 
%This choice actually favors the observation of superradiance~\cite{Wiegner2015}. 
In this way, only minor spatial modulations of the spontaneous emission rate have been observed, attributed to multiple scattering~\cite{Araujo2016} or to linear dispersion~\cite{Kuraptsev2017}.

%In this context, subwavelength systems and atoms in cavities exhibit the strongest modification~\cite{Dicke1954,gross1982},  , where dipole-dipole interactions are strongest, have their emission largely dominated by superradiance, as first described by Dicke in his seminal work~\cite{Dicke1954,gross1982}. This picture of a large collection of emitters which decays through a series of symmetric superradiant states was ...., which stems from the fact that the weak decoherence which characterizes subradiant states also comes with a weak coupling to an external pump****. Furthermore, we note that although conditional measurements guarantees that the system is cast into entangled states[REF 12 - 2atom], which are the states originally envisioned by Dicke~\cite{Dicke1954}, it actually forces the system into the 

%since very close in-phase dipoles 

% {\it Here, however, subradiant decay can be observed only if the separation between the particles remains larger than half the transition wavelength -- in which case the modification of the spontaneous emission rate is only moderate.}
In this work we show that a pronounced spatial modulation of the spontaneous emission rate can be obtained  by combining dipole-dipole interactions among the emitters with conditional measurements of the scattered photons. We consider the simplest configuration where this phenomenon is observable, namely three atoms, where two of them are separated at sub-wavelength distance from each other while the third one is located several wavelengths apart, all entangled via conditional photon measurement \cite{Cabrillo1999,Skornia2001,Plenio2003,Moehring2007,Hofmann2012,Slodivcka2013,Bernien2013,Delteil2016,Richter2022}. In this configuration, the  strong dipole-dipole interaction between the adjacent atoms leads to a marked modification of the spontaneous decay rate \cite{Dicke1954}; the latter is, however, almost isotropic due to the close separation of the two emitters. The decay rate becomes modulated in space only due to the presence of the remote atom, entangled with the other two atoms.
% we demonstrate that by adding a third non-interacting atom at a  large distance of several wavelength with respect to the other two atoms, this limitation can be bypassed. 
By measuring Glauber's third-order photon correlation function where two photons are initially recorded in given directions, all three atoms become entangled  giving rise to spatially varying quantum interferences among the different decay channels of the three atoms. In this case a modulated spatial pattern for the emission rate of the last photon is obtained, displaying both superradiant and subradiant decay. Note that a two-atom system cannot support both a strong modification of the spontaneous emission rate (obtained for close atoms) and a spatial modulation (accessible only with remote atoms). Our work thus paves the way for engineering  the photon emission rate by taking advantage of both dipole-dipole interactions and distant atom-atom correlations.
% By measuring Glauber's third-order intensity correlation function $G^{(3)}(\bm{r}_1,\bm{r}_2,\bm{r}_3, t_1,t_2,t_3)$, which describes the conditional probability of measuring a photon at position $\bm{r}_3$ at time $t_3$, given that two photons were measured at $(\bm{r}_2, t_2)$ and $(\bm{r}_1, t_1$), pronounced spatial and temporal modifications of the spontaneous emission rate for the last released photon are achieved leading to pronounced super- and subradiant emission rates.
%A typical spatio-temporal interference pattern is shown in Fig.~\ref{fig:geometry} (b). 
%The third-order correlation function differs significantly from the usual atomic intensity pattern (first-order correlation function) as the temporal course of the latter would be independent of the detection direction.

\begin{figure*}
	\centering
	%\input{tikz/setup-geometry.tex}
				% left lower right upper
	\includegraphics[width=\textwidth]{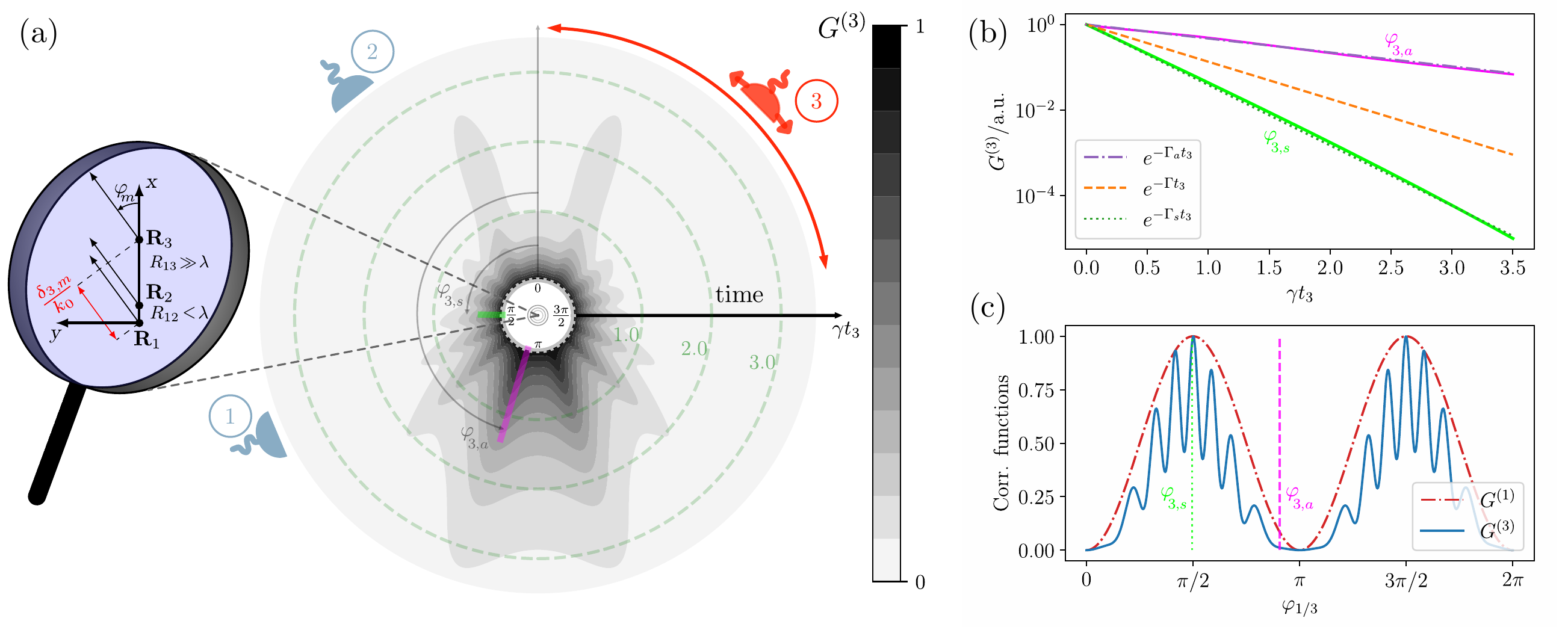}
	\caption{(a) Three identical two-level atoms placed at $\bm{R}_1 = \bm{0}$, $\bm{R}_2 =\lambda/3 \hat{\bm{x}}$ and $\bm{R}_3=4\lambda \hat{\bm{x}}$ along the $x$-axis; with $R_{12} < \lambda$, the first two atoms are subject to the dipole-dipole interaction, while the third atom, with $R_{23}\gg \lambda$, is not. %the spontaneously emitted photons are recorded in the far field by three detectors at $\bm{r}_m$, $m\in \{1,2,3\}$. 
	Starting from the fully excited state, the three-atom system is projected onto an entangled state via the measurement of two photons at space-time points $(\varphi_1=2\pi/3, t_1=0)$ and $(\varphi_2=\pi/4.4, t_2=0)$  (blue detectors). The signal of the third detector (red) then corresponds to the probability to record the last photon at space-time point $(\varphi_3, t_3)$,  collectively scattered by the entangled atomic ensemble. The contour plot shows the third-order photon correlation function $G^{(3)}(\varphi_{3},t_3)$, %as a function of the third detection angle $\varphi_3$ and time $\gamma t_3$ (radial component), 
	normalized for each direction by its initial value; angles $\varphi_{3,s}\approx 1.56$ ($\varphi_{3,a}\approx 2.85$) indicate the directions at which an effective superradiant (subradiant) decay is observed. % The time $\gamma t_3$ is plotted along the radial component, whereas the detection angle $\varphi_3$ refers to the actual angle along the disc.	One finds a rich spatial interference pattern and a modified decay rate depending on the direction of observation. 
	 (b) Time evolution of $G^{(3)}(\varphi_{3},t_3)$, normalized by its value at time zero, for the two directions $\varphi_{3,s}$ (thick lime line) and $\varphi_{3,a}$ (thin magenta line), where the decay rate is approximately given by the symmetric $\Gamma_s = 2(\gamma+\Delta\gamma)$ and antisymmetric $\Gamma_a = 2(\gamma-\Delta\gamma)$ decay rate, respectively; the dash-dotted purple, dashed orange and dotted green lines are exponential curves displaying the antisymmetric, single atom and symmetric decay rates, respectively. %\textcolor{magenta}{(1) let us add a curve for a direction with single-atom decay rate? (2) We should pick a single notation: $(\gamma+\Delta\gamma)$ or $\gamma_S$. Idem for antisym. I suggest choosing $\gamma_S$, more compact, and more consistent with the discourse on sym/antisym modes.}
% 	 along the lime and magenta lines. At these specific angles, the correlation function decays with effective decay rates approximately given by the symmetric or antisymmetric decay rates $\gamma \pm \Delta \gamma$. The purple, green and orange dashed lines correspond to the analytic symmetric, antisymmetric and single atom decay, respectively.
    (c) Third-order correlation function $G^{(3)}(\varphi_{3},t_3=0)$ (blue solid curve) and first-order correlation function  $G^{(1)}(\varphi_{1},t_1=0)$  (red dash-dotted curve) at  initial times; 
    the entanglement of the three-atom system created by the measurement of the first two photons leads to a strong modulation in space of $G^{(3)}(\varphi_{3},t_3=0)$, whereas $G^{(1)}(\varphi_{1},t_1=0)$ displays only the dipole radiation pattern; the direction at which an effective superradiant (subradiant) decay is observed is indicated by the dotted lime (dashed magenta) line. %Atoms are positioned at $\bm{R}_1 = \bm{0}$, $\bm{R}_2 =\lambda/3 \hat{\bm{x}}$ and $\bm{R}_3=4\lambda \hat{\bm{x}}$, and the first two detection angles are $\varphi_1=2\pi/3$ and $\varphi_2=\pi/4.4$ for all three panels.
    }
	\label{fig:geometry}
\end{figure*}

We start by investigating the emission dynamics by use of the master equation for three identical two-level atoms. To demonstrate the effect, it is sufficient to restrict the analysis  to three atoms located at positions $\bm{R}_\mu=(x_\mu,y_\mu)$, $\mu \in \lbrace 1, 2, 3 \rbrace$ within the $xy$-plane, as illustrated in Fig.~\ref{fig:geometry}(a). Here, two atoms are assumed to be adjacent with sub-wavelength separation such that they are subject to the dipole-dipole interaction while the third atom is located at a distance of several wavelengths from the other two emitters, so that the light-mediated interaction with the first two atoms can be neglected. The state of the atomic system is conveniently expressed using the collective Dicke-basis for the first two atoms
\begin{align}
	&\ket{E}=\ket{e,e}\,,\quad \ket{G} = \ket{g,g}\,,\nonumber\\
	&\ket{S}=\frac{1}{\sqrt{2}}(\ket{e,g}+\ket{g,e})\,,\\
	&\ket{A}=\frac{1}{\sqrt{2}}(\ket{e,g}-\ket{g,e})\,,\nonumber
\end{align}
and the bare atomic basis with excited state $\ket{e}$ and ground state $\ket{g}$ for the remaining third atom. In this case, the master equation for the three-atom density matrix $\op{\rho}$ reads~\cite{Agarwal1974,FicekSwain2005}
\begin{equation}
\begin{split}
&\partial_t \op{\rho} = - i \omega_0 \sum_{\mu=1}^3 \Big[ \op{S}_{z}^{(\mu)}, \op{\rho} \Big] + i  \sum_{\mathclap{\substack{\mu,\nu=1 \\ \mu \neq \nu}}}^2 \Delta \Omega \Big[ \op{S}_{+}^{(\mu)}\op{S}_{-}^{(\nu)}, \op{\rho} \Big] \\
&- \sum_{\mu=1}^3 \gamma \big( \op{S}_{+}^{(\mu)}\op{S}_{-}^{(\mu)} \op{\rho} - 2 \op{S}_{-}^{(\mu)} \op{\rho} \op{S}_{+}^{(\mu)} + \op{\rho} \op{S}_{+}^{(\mu)}\op{S}_{-}^{(\mu)}  \big) \\
&- \sum_{\mathclap{\substack{\mu,\nu=1 \\ \mu \neq \nu}}}^2 \Delta \gamma \big( \op{S}_{+}^{(\mu)}\op{S}_{-}^{(\nu)} \op{\rho} - 2 \op{S}_{-}^{(\nu)} \op{\rho} \op{S}_{+}^{(\mu)} + \op{\rho} \op{S}_{+}^{(\mu)}\op{S}_{-}^{(\nu)}  \big) \, .
\label{eq:maeq}
\end{split}
\end{equation} 
Here, $\op{S}_{+}^{(\mu)}$ ($\op{S}_{-}^{(\mu)}$) denotes the raising (lowering) operator of the $\mu$th atom and $\op{S}_{z}^{(\mu)} = \frac{1}{2}(\op{S}_{+}^{(\mu)}\op{S}_{-}^{(\mu)} - \op{S}_{-}^{(\mu)}\op{S}_{+}^{(\mu)})$. Moreover, $\omega_0=k_0 c=2\pi c/\lambda$ stands for the atomic transition frequency, $\Gamma=2\gamma$ the single atom decay rate, whereas the coupling parameters $\Delta\gamma$ and $\Delta\Omega$ account for the dipole-dipole interaction between the first two atoms 
\begin{equation}
\begin{split}
\Delta \Omega - i \Delta \gamma &= \frac{3}{2} \gamma e^{-i k_0 R_{12}} \left[ \frac{1-\cos^2\psi}{k_0 R_{12}} \right.\\
&\left. \hspace*{-.8cm}- [1-3\cos^2\psi] \left( \frac{i}{(k_0 R_{12})^2} +\frac{1}{(k_0 R_{12})^3}  \right) \right] \,
\end{split}
\label{eq:deltagammaomega}
\end{equation}
with $R_{12}=|\bm{R}_{12}|=|\bm{R}_1-\bm{R}_2|$ and $\psi$ the angle between the atomic dipole moment $\bm{d}$ and $\bm{R}_{12}$.\\
\indent We assume the system to be initially in the fully excited state $\ket{E,e}$. To calculate the third-order photon correlation function, we assume that two detectors at $\bm{r}_1$ and $\bm{r}_2$ record two photons spontaneously emitted by the three-atom system at time $t_1=t_2=0$, whereas the third photon is measured at position $\bm{r}_3$ at time $t_3$. The detection of a photon corresponds to the annihilation of the light particle, described by the positive frequency part of the electric field operator $\op{\bm{E}}_m^{(+)}$ (with $\op{\bm{E}}_m^{(-)}=[\op{\bm{E}}_m^{(+)}]^\dagger$). Calling $\hat{\bm{r}}_m=\bm{r}_m/r_m= \hat{\bm{x}}\cos\varphi_m  +  \hat{\bm{y}}\sin\varphi_m$ the direction of the detector in the far field we can write
\begin{equation}
\op{\bm{E}}^{(+)}_m \propto \hat{\bm{r}}_m\times(\hat{\bm{r}}_m\times \bm{d})\cdot \sum_{\mu=1}^{3} e^{i \delta_{\mu,m}} \op{S}_{-}^{(\mu)} \,,
\label{eq:opD}
\end{equation}
where the cross product yields the usual dipole radiation pattern. In addition, the phase
\begin{equation}
\begin{split}
\delta_{\mu,m} &= -k_0{\bm{R}_\mu} %\frac{k_0 {\bm{r}_m}}{\abs{{\bm{r}_m}}}
\cdot \hat{\bm{r}}_m\\
&= -k_0 [x_\mu \cos\varphi_m  + y_\mu \sin\varphi_m]
\end{split}
\label{eq:geo_phases}
\end{equation}
accounts for the relative geometric phase (or optical path) accumulated by a photon when propagating from the emitter at $\bm{R}_\mu$ to the detector at $\bm{r}_m$ relative to a photon emitted at the origin
%, as compared to photons emitted by atom $1$ at the origin
[see Fig.~\ref{fig:geometry} (a)].% REPETITION, SINCE ALREADY INTRODUCED BEFORE (4), and $\varphi_m$ is the azimuthal angle of the $m$th detector.\\
\indent Considering the fully excited state Glauber's third-order photon correlation function, i.e., the conditional probability to record a photon at the third detector at space-time point $(\bm{r}_3, t_3)$ given that two photons have been measured at $(\bm{r}_2, 0)$ and $(\bm{r}_1, 0)$, we find ~\cite{Oppel2014,Wiegner2015,SimonLukas2020}
\begin{equation}
\begin{split}
G^{(3)}&(\bm{r}_1,\bm{r}_2,\bm{r}_3;0,0,t_3) \\
&= \langle \op{\bm{E}}^{(-)}_1 \op{\bm{E}}^{(-)}_2 \op{\bm{E}}^{(-)}_3 \op{\bm{E}}^{(+)}_3 \op{\bm{E}}^{(+)}_2 \op{\bm{E}}^{(+)}_1 \rangle_{\ketbra{E,e}{E,e}}\,. \\
\end{split}
\label{eq:G3eee}
\end{equation}
\noindent Without loss of generality, we can set $\bm{R}_1=(0,0)$ leading to \cite{SimonLukas2020}
\begin{widetext}
	\begin{equation}
	\begin{split}
	G^{(3)}(\bm{r}_3,t_3) &\propto \sin^2(\alpha) \left[\abs{c_{Ge}}^2 e^{- 2 \gamma t_3} \,+\, 2 \abs{c_{Sg}}^2 e^{-2 (\gamma + \Delta \gamma) t_3} \cos^2\left(\delta_{2,3}/2\right) \,+\, 2 \abs{c_{Ag}}^2 e^{-2 (\gamma - \Delta \gamma) t_3} \sin^2\left( \delta_{2,3}/2 \right)\right. \\
	&\left.+ 2 \abs{c_{Sg}} \abs{c_{Ag}} e^{- 2 \gamma t_3} \sin\left( \delta_{2,3} \right) \sin\left(\varphi_{Ag} - \varphi_{Sg} - 2 \, \Delta \Omega \, t_3 \right)\right. \\
	&\left.+ 2 \sqrt{2} \abs{c_{Sg}} \abs{c_{Ge}} e^{- (2 \gamma + \Delta \gamma) t_3} \cos\left( \delta_{2,3}/2 \right) \cos\left(\varphi_{Sg} + \delta_{2,3}/2 - \delta_{3,3} + \Delta \Omega \, t_3 \right)\right. \\
	&\left.+ 2 \sqrt{2} \abs{c_{Ag}} \abs{c_{Ge}} e^{- (2 \gamma - \Delta \gamma) t_3} \sin\left( \delta_{2,3}/2 \right) \sin\left(\varphi_{Ag} + \delta_{2,3}/2- \delta_{3,3} - \Delta \Omega \, t_3 \right)\right] \, ,
	\end{split}
	\label{eq:G3sol}
	\end{equation}
\end{widetext}
where the coefficients $c_{Ge}$, $c_{Sg}$ and $c_{Ag}$ read 
\begin{equation}
\begin{split}
&c_{Ge} = \left( e^{i \delta_{2,1}} + e^{i \delta_{2,2}} \right) \,, \\
&c_{Sg} = \frac{1}{\sqrt{2}} \left( e^{i (\delta_{3,1}+\delta_{2,2})}   +  e^{i (\delta_{2,1}+\delta_{3,2}) }   +  e^{i \delta_{3,1}}  +  e^{i \delta_{3,2}}  \right) \,, \\
&c_{Ag} = \frac{1}{\sqrt{2}} \left( e^{i (\delta_{3,1}+\delta_{2,2})}   +  e^{i (\delta_{2,1}+\delta_{3,2}) }   -  e^{i \delta_{3,1}}  -  e^{i \delta_{3,2}}  \right)  \,,
\end{split}
\label{eq:coeff}
\end{equation} 
and the phases associated with the complex coefficients $c_{Sg}$ and $c_{Ag}$ are given by $\varphi_{Sg/Ag} = \text{Arg}( \frac{c_{Sg/Ag}}{c_{Ge}})$. In Eq.~(\ref{eq:G3sol}), the $\sin^2(\alpha)$ term accounts for the dipole radiation pattern, with $\alpha$ the angle between the direction of the last detector $\hat{\bm{r}}_3$ and the dipole moment $\bm{d}$ of the atoms. %Further, the indices $S$, $A$ and $G$ refer to the usual symmetric, antisymmetric and double ground state of the Dicke basis for two atoms.

In what follows, we place the three atoms along the $x$-axis and assume the dipole moment to be parallel to this axis, $\bm{d}=d\hat{\bm{x}}$, such that $\alpha=\varphi_3$ [see Fig.~\ref{fig:geometry} (a)]. The spatial and temporal behavior of $G^{(3)}(\bm{r}_3,t_3)$ is thus tuned by the six geometrical phases $\delta_{\mu,m}$, $\mu \in \{2,3\}$ [with  $\bm{R}_1=(0,0)$] and $m\in\{1,2,3\}$. This large parameter space results in a great variety of atom geometries, state preparation and detection setups, each with different outcome for $G^{(3)}(\bm{r}_3,t_3)$. For example, the specific cases $|c_{Sg}|=0$ or $|c_{Ag}|=0$ reveal regimes in which $G^{(3)}(t_3)$ strongly deviates from a true exponential decay and rather presents either a global maximum or a true root $G^{(3)}(t_3)=0$ at finite time $t_3>0$, corresponding to birth and death of spontaneous emission, respectively~\cite{SimonLukas2020}.

\begin{figure}
	\hspace*{-.4cm}
	\includegraphics[width=.4\textwidth]{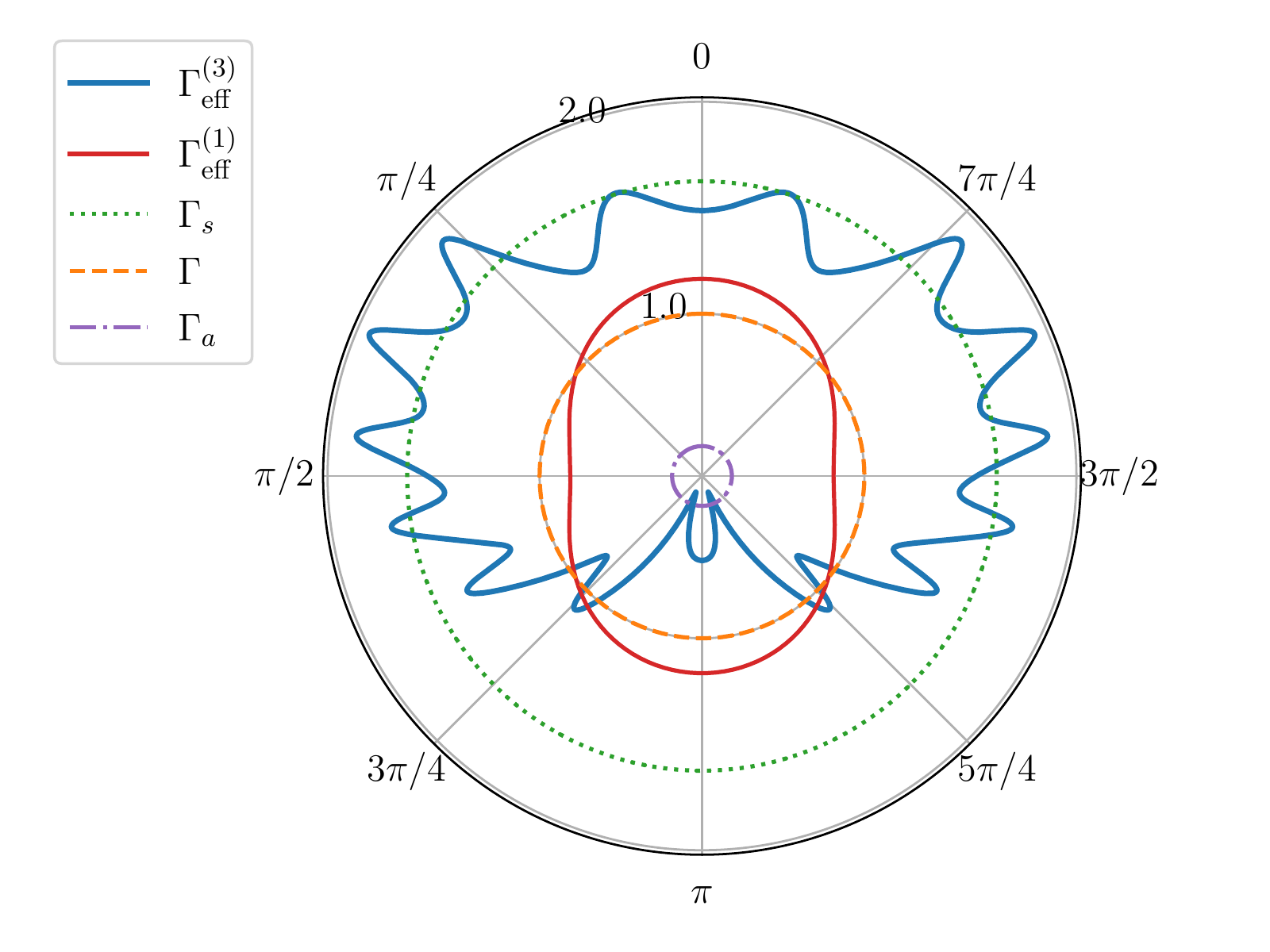}
	\caption{Effective decay rate $\Gamma^{(3)}_\mathrm{eff}$ of the third-order photon correlation function $G^{(3)}(\varphi_{3},t_3)$ as a function of the direction of observation (thick blue solid curve), together with the effective decay rate $\Gamma^{(1)}_\mathrm{eff}$ of the intensity measurement $G^{(1)}(\varphi_{1},t_1)$ (red solid line), the symmetric decay rate $\Gamma_s=2(\gamma+\Delta\gamma)$ (green dotted line), the single-atom decay rate $\Gamma$ (orange dashed line), and the antisymmetric decay rate $\Gamma_a=2(\gamma-\Delta\gamma)$ (purple dash-dotted line) for the same setup as in Fig.~\ref{fig:geometry}. The effective decay rates, all presented in units of $\gamma$, are calculated by an exponential fit of the correlation functions in the time interval $\gamma t \in [0,0.5]$. %We find several different decay rates for the third-order intensity correlation function depending on the detection angle. In particular, the $G^{(3)}$ measurement allows for the observation of subradiant decay in contrast to a usual $G^{(1)}$ measurement.
	}
	\label{fig:dec_rates}
\end{figure}

If the three atoms are placed at $\bm{R}_1 = \bm{0}$, $\bm{R}_2 =\lambda/3 \hat{\bm{x}}$ and $\bm{R}_3=4\lambda \hat{\bm{x}}$ and the first two photon detection events at time zero occur in space at $\varphi_1=2\pi/3$ and $\varphi_2=\pi/4.4$, then Glauber's third-order correlation function $G^{(3)}(\bm{r}_3,t_3)$, normalized for each direction $\varphi_3$ by its value at time zero, takes the form shown in Fig.~\ref{fig:geometry} (a). In that case both super- and subradiant decay can be observed along certain directions, as illustrated in Fig.~\ref{fig:geometry}~(b). 
Note, however, that in general the emission dynamics in each direction results from the sum of different modes [see Eq.~\eqref{eq:G3sol}].

This variety of decay rates in space comes with a rich spatial interference pattern of $G^{(3)}(\bm{r}_3,t_3=0)$ exhibiting a series of fringes [Fig.~\ref{fig:geometry} (c)]. The pattern stems from the quantum interference of the emission probability of the remote atom with that of the other two emitters, with the latter two being too close to produce a spatial modulation on their own. The number of fringes is directly related to the distance of the third atom to the other two atoms, i.e., putting it farther away will increase that number.

We highlight that this great variety of modified spontaneous emission in both space and time results from the combined action of dipole-dipole interaction and conditional measurements leading to entanglement of the sources. Indeed, although one may argue that interference of the light fields among the three emitters occurs independently of the conditional measurement intrinsic to $G^{(3)}(\bm{r}_3,t_3)$, leaving aside the conditional measurements alters the emission pattern drastically. This is due to the fact that the states responsible for the light emission are very different in both cases. While only an isotropic emission pattern is obtained for $G^{(1)}(\bm{r}_1,t_1=0)$ starting from the state $\ket{E,e}$ [up to the dipole radiation pattern $\sin^2(\alpha)$], $G^{(3)}(\bm{r}_3,t_3=0)$ displays strong spatial modulations.
%also at later times the interference between the light from the two close atoms with that of the remote atom produces only a modest spatial modulation. 
In Fig.~\ref{fig:geometry} (c), the emission patterns obtained from Glauber's first- and third-order correlation functions are presented. 
%Indeed, only a weak spatial variation of the emission pattern is observed, i.e.,
As can be seen, the superposition of the light fields from the two close atoms with the one of the remote atom fails to produce the intricate fringe pattern produced by the three-atom system entangled via the conditional measurements \cite{Thiel2007}.

Moreover, a careful analysis of the emission dynamics reveals how the entanglement of the  atoms affects also the temporal emission properties of the three-atom system. In Fig.~\ref{fig:dec_rates} (thick blue solid curve) the effective decay rates of $G^{(3)}(\bm{r}_3,t_3)$ are computed for different directions of observation for the same conditional measurement configuration as in Fig.~\ref{fig:geometry}. The decay rates are obtained by fitting exponentially the radiation dynamics of $G^{(3)}(\bm{r}_3,t_3)$ in the time interval $\gamma t\in [0, 0.5]$. Note that the emission from the entangled states associated with a $G^{(3)}$ measurement allows all three modes (symmetric, antisymmetric and single atom) to contribute to the temporal emission behavior, thus producing the intricate pattern of effective decay rates displayed in Figs.~\ref{fig:geometry}~(a) and \ref{fig:dec_rates}. The latter is in strong contrast to the pattern obtained by a direct measurement of the decaying intensity $G^{(1)}$, i.e., obtained without conditional measurement, leading merely to a weak modulation of the decay rate in space as shown in Fig.~\ref{fig:dec_rates} (red solid curve). This modulation results only from the interference between the emission of the first two atoms. 

We note that while the dipole-dipole interaction between two atoms allows for modifications of the decay rate which are either larger or smaller than the single-atom decay rate [i.e., the signatures of super- and subradiance as in Eq.~(\ref{eq:deltagammaomega})], $G^{(3)}(\bm{r}_3,t_3)$ displays a multitude of directions with faster-than-symmetric and slower-than-antisymmetric decay rates (see Fig.~\ref{fig:dec_rates}). Indeed, the coherent part of the dipole-dipole interaction leads to frequency shifts of the collective modes and eventually mode beating since the different modes contributing to the radiation in Eq.~\eqref{eq:G3sol} compete. The obtained oscillations result in an increase or decrease of the decay rates at initial times~\cite{SimonLukas2020}, surpassing the symmetric and antisymmetric decay rate in certain directions.

In conclusion, we demonstrated how conditional measurements combined with dipole-dipole interactions enable manipulation and even engineering of the spontaneous emission behavior, both in space and time, of an atomic ensemble. Dipole-dipole interactions are strongest for subwavelength samples~\cite{Dicke1954}, from which, however, no appreciable interference pattern is obtained. Imposing correlations between remote atoms by conditional measurements allows one to bypass this restriction of short interatomic distances; this becomes an option because the generation of entanglement via conditional measurements is possible even for remote atoms~\cite{Thiel2007,Bastin2009}. In this work, we have shown how to combine these two atom-correlating processes within the simplest configuration, i.e., using a pair of close atoms correlated to a remote one by conditional measurement.

The marked quantum interference resulting from the conditional measurements reveals directions along which subradiant decay dominates the emission, from the earliest moment and at odds from the two-atom case~\cite{SimonLukas2020}. 
%This mechanism is particularly promising since subradiant states are notoriously difficult to address~\cite{Kalachev2006,Kalachev2007,Scully2015,Facchinetti2016,Cipris2021,Ferioli2021}.
However, 
%we restate that the emission dynamics in each direction results in general from the sum of different modes. Moreover, 
it should be mentioned that in the three-atom case, congruent to the two-atom case, the subradiant mode scales in the small-distance limit as $c_{Ag}\propto R_{12}/\lambda$. Thus increasing the lifetime of the subradiant mode goes along  with a decrease in the corresponding population. In the future, we will investigate how subsequent conditional measurements in space {\it and} time will modify the collective decay and how the behavior scales with increasing number of emitters. This work shows that collective spontaneous emission is a rich line of research with unexpected outcomes beyond the canonical two-atom case, even if only a single atom is added to the system.
\begin{acknowledgments}
	M.B. and L.G. gratefully acknowledge funding and support by the International Max Planck Research School - Physics of Light. R.B. and J.v.Z. gratefully acknowledge funding and support by the  Bavarian Academic Center for Latin America (BAYLAT). This work was funded by the Deutsche Forschungsgemeinschaft (DFG, German Research Foundation) -- Project-ID 429529648 -- TRR 306 QuCoLiMa ("Quantum Cooperativity of Light and Matter''). R.B. is supported by the S\~ao Paulo Research Foundation (FAPESP, Grants Nos. 2018/15554-5 and 2019/13143-0) and by the Brazilian National Council for Scientific and Technological Development (CNPq, Grants Nos. 313886/2020-2 and 409946/2018-4). 
\end{acknowledgments}

\vspace*{-.4cm}
\vfill

\pagebreak

\bibliography{library}
 
% \vspace*{2cm} %%refs in one column without main text
% 
% 
% \pagebreak
% 
% \appendix*
% 
% 

\end{document}